\documentstyle[sprocl]{article}

\def\Journal#1#2#3#4{{#1} {\bf #2}, #3 (#4)}

\def\RMP{\em Rev. Mod. Phys.}
\def\NPB{{\em Nucl. Phys.} B}
\def\APPB{{\em Acta Phys. Polonica} B}
\def\NPA{{\em Nucl. Phys.} A}
\def\YAF{\em Yad. Phys.}
\def\PLB{{\em Phys. Lett.} B}
\def\PL{\em Phys. Lett.}
\def\PRP{\em Phys. Rep.}
\def\UFN{\em Uspekhi Fiz. Nauk}
\def\RPP{\em Rep. Prog. Phys.}
\def\PRL{\em Phys. Rev. Lett.}
\def\PRD{{\em Phys. Rev.} D}
\def\PRC{{\em Phys. Rev.} C}
\def\PR{\em Phys. Rev.}
\def\IJP{\em Indian J. Phys.}

\def\ZPA{{\em Z. Phys.} A}
\def\PRSLA{{\em Proc. R. Soc. London} A}
\def\PTPS{\em Prog. Theor. Phys. Suppl.}

\def\be{\begin{equation}}
\def\ee{\end{equation}}
\def\bea{\begin{eqnarray}}
\def\eea{\end{eqnarray}}

\begin{document}
\title{THE OZI RULE VIOLATION IN $N \bar N$ ANNIHILATION AT REST
AND THE SKYRME MODEL}
\author{ Z.K. SILAGADZE }
\address{Budker Institute of Nuclear Physics \\ 630 090,
Novosibirsk, Russia
\\E-mail: silagadze@inp.nsk.su}
\maketitle
\abstracts{
The question is raised if a large violation of the OZI rule,
recently observed in many channels in $N \bar N$ annihilation
at rest, can be explained in the framework of the Skyrme model.
}

\section{Introduction}
According to the OZI-rule \cite{1}, $\phi$-meson production in
$N \bar N$ annihilation can happen only due to its nonstrange
quark content and should be significantly suppressed \cite{2} :
\begin{eqnarray}
R(X)=\frac{BR(N \bar N \to \phi X)}{BR(N \bar N \to \omega X)}
\sim tg^2 \delta = (2.5 \pm 1.5) \cdot 10^{-3} \; \; ,
\nonumber
\end{eqnarray} \noindent
where $X$ stands for the accompanying particle(s) and
$\delta \approx 4^o$ is a small deviation from the ideal $\omega -
\phi $ mixing angle.

This prediction is not confirmed by recent high-statistics experiments
at LEAR. A large excesses over the naive OZI expectations were reported
in the $\phi$-meson production in $N \bar N$ annihilation at rest by
ASTERIX, Crystal Barrel and OBELIX collaborations \cite{3,4,5}. In the
most distinguished channels the OZI rule violation reaches about two
orders of magnitude:
\begin{eqnarray}
R(\pi _0)=0.14 \pm 0.04   \; \; , \; \;
R(\pi _-)=0.16 \pm 0.04   \; \; , \; \;
R(\gamma)=0.33 \pm 0.15   \; .
\nonumber
\end{eqnarray} \noindent
The weaker violation of the quark-line rule is also observed in many
other channels.

The $\phi$-meson production in nucleon antinucleon annihilation at rest
reveals the following features \cite{6} :
\begin{itemize}
\item There are strong enhancement over the naive OZI prediction
for the channels $\phi \pi$, $\phi \gamma$. In the channels
$\phi \rho , \phi \omega , \phi \pi \pi$ there are by one order of
magnitude smaller enhancement.
\item There is no enhancement at all in the $\phi \eta$ channel.
\item For nucleon antinucleon annihilation at rest the OZI rule violation
is much more stronger than for $\pi p$ or $p p$ scattering and higher
energy $p \bar p$ annihilation.
\item The large enhancement of $\phi \pi$ appears to be due to S-wave
annihilation, P-wave annihilation exhibiting no large deviation from
the OZI prediction.
\end{itemize}

Several models were suggested to explain the large $\phi$ production
rate in $N \bar N$ annihilation at rest, neither of them being completely
  successful in incorporating the above mentioned features of the
annihilation.

The most popular explanation of the OZI-rule violation assumes the
existence of the strange sea quarks admixture in proton \cite{6,7}.
There are indications also from other areas of the elementary particle
physics that the nucleon wave function contains some amount of $s \bar s$
pairs already in the non-perturbative regime at large distances
\cite{4,7,8}. In this picture, a large OZI-rule violation is interpreted
in terms of the "shake-out" and "rearrangement" of an $s \bar s$ component
of the nucleon wave function \cite{6}. Why different channels are so
drastically unequally effected by this strange sea, remains a mystery.

Another model involves an exotic four-quark resonance, presumably
dubious $C(1480)$ meson, to explain an enhancement in the S-wave
$\phi \pi$ production and lack of enhancement in the P-wave channel
\cite{9}. However even more stronger OZI-rule violation in the $\phi
\gamma$ channel might not be due to this mechanism, because $\phi
\gamma$ has positive $C$-parity in contrast to $\phi \pi^0$ and can not
be produced via intermediate crypto-exotic $1^{--}$ resonance.

One more possibility to overcome the quark line rule is to assume
that $\phi$ meson is produced due to the final state interaction of
kaons formed in the OZI allowed process $p \bar p \to K^* \bar K \to
K \bar K \pi \to \phi \pi$ \cite{10}. But this rescattering model can
not explain the absence of OZI rule violation in the $\pi \pi \phi$
channel despite the fact that $K^* \bar K^*$ final state is even more
copious than $K^* \bar K$ in $N \bar N$ annihilation \cite{5}.
The concrete calculations in this model \cite{10,11} also don't reproduce
experimental data very well, giving more moderate OZI rule violation than
observed.

Theoretically the quark line rule is justified by the $\frac{1}{N_c}$
expansion which shows that non-valence $q \bar q$ pairs are suppressed
in mesons \cite{12}. But for baryons, owing the fact that they contain
$N_c$ valence quarks, non-valence $q \bar q$ pairs is expected to be
$O(N_c)$ more important than in mesons \cite{7}. So for baryons the OZI
rule has no theoretical background even in large $N_c$ limit \cite{7}.

On the other hand, the structure of QCD simplifies in the large $N_c$
limit, where it is equivalent to an effective theory of weakly
interacting mesons, as was shown by Witten \cite{12}. We expect that
this simplified effective theory also reveals the OZI rule violation
for baryons and thus is a good starting point to study the phenomenon.

\section{The Skyrme model}
In the low energy limit, the (approximate) $SU(3) \times SU(3)$ chiral
symmetry of QCD is spontaneously broken and the $\gamma _5$-phase of quark
fields have nonzero vacuum expectation value. Let us rewrite the QCD
lagrangian separating the vacuum chiral phase, which can be considered
as an external field
\begin{eqnarray}
{\cal L} & = & \bar \psi \exp {\{-\frac{i}{2f_\pi}\gamma_5 \phi^a
\lambda^a \}}i\hat {\partial} \exp {\{\frac{i}{2f_\pi}\gamma_5
\phi^a \lambda^a\}} \psi + \cdots  \\ \nonumber
& = & \bar \psi _R i (\hat {\partial}+\hat {A}_R)\psi_R +
 \bar \psi _L i (\hat {\partial}+\hat {A}_L)\psi_L + \cdots \; ,
\label{eq1} \end{eqnarray} \noindent
where $A_{R\mu}=W^+\partial_\mu W, A_{L\mu}=W\partial_\mu W ^+ $ and
$W=\exp ({\frac{i}{2f_\pi}\phi^a \lambda^a})$.

In fact, pseudoscalar mesons are excitations of just this $\gamma _5$-phase
above the vacuum. More precisely, because mesons are quark-antiquark bound
states, they should be associated with the chiral phase of the quark
bilocal $\bar q_L q_R$, that is with $U=\exp {(\frac{i}{f_\pi}\phi^a
\lambda^a)} = W^2$.

Integrating the quark (and gluon) degrees of freedom from (1), we end with
the effective theory of the $U(x)$ field:
\begin{eqnarray}
{\cal L}_{eff.} = \frac{f_\pi^2}{4} Sp (\partial_\mu U)(\partial^\mu U^+) +
\cdots
\label{eq2} \end{eqnarray}
We can expect that in the low energy limit the other terms in (2),
containing more derivatives, are less significant when the first one.

But (2) is a purely meson theory. Where are baryons? The crucial idea of
T.~Skyrme \cite{13}, revived in the context of QCD by E.~Witten \cite{14},
is that we can still describe baryons by (2), considering them as
vortexes of the mesonic "liquid".

Let us establish the most striking features of the Skyrme model, that
these vortexes can carry a nonzero baryon number and a half-integer spin.

Maybe a nonzero baryon number, localized at Skyrmion, will appear not so
strange if we remember about the negative energy Dirac's sea of quarks.
When the emergence of the baryon number can be considered as a "vacuum
polarization" effect in the strong meson field of Skyrmion \cite{15}.

Because of the chiral anomaly, from (1) we have \cite{16}
\begin{eqnarray}
\partial_\mu \bar \psi_R \gamma^\mu \psi_R & = & \frac{-1}{32\pi^2}
Sp F_{\mu \nu}^R \tilde F_R^{\mu \nu} = \frac{-1}{8\pi^2}
\partial^\mu \varepsilon_{\mu \nu \lambda \sigma} Sp (A_R^\nu \partial
^\lambda A_R^\sigma + \frac{2}{3} A_R^\nu A_R^\lambda A_R^\sigma )
\nonumber \\
\partial_\mu \bar \psi_L \gamma^\mu \psi_L & = & \frac{1}{32\pi^2}
Sp F_{\mu \nu}^L \tilde F_L^{\mu \nu} = \frac{1}{8\pi^2}
\partial^\mu \varepsilon_{\mu \nu \lambda \sigma} Sp (A_L^\nu \partial
^\lambda A_L^\sigma + \frac{2}{3} A_L^\nu A_L^\lambda A_L^\sigma )
\; . \nonumber \end{eqnarray} \noindent
But $A_\mu^R=W^+\partial_\mu W$ and $A_\mu^L=W\partial_\mu W^+$ lead to
$$\varepsilon_{\mu \nu \lambda \sigma} \partial^\lambda A_R^\sigma=
- \varepsilon_{\mu \nu \lambda \sigma} A_R^\lambda A_R^\sigma \; \; ,
\; \;
\varepsilon_{\mu \nu \lambda \sigma} \partial^\lambda A_L^\sigma=
- \varepsilon_{\mu \nu \lambda \sigma} A_L^\lambda A_L^\sigma $$
\noindent and $Sp(A_L^\nu A_L^\lambda A_L^\sigma)=-
Sp(A_R^\nu A_R^\lambda A_R^\sigma)$. Thus
\begin{eqnarray}
\partial_\mu \bar \psi_R \gamma^\mu \psi_R =
\partial_\mu \bar \psi_L \gamma^\mu \psi_L =
\frac{1}{24\pi^2}
\partial^\mu \varepsilon_{\mu \nu \lambda \sigma} Sp ( A_R^\nu
A_R^\lambda A_R^\sigma )  \; .
\label{eq3} \end{eqnarray}
For the baryon current $B_\mu = N_c \cdot \frac{1}{N_c} \bar \psi
\gamma_\mu \psi = \bar \psi \gamma_\mu \psi$ (the first $N_c$ factor is
due to color), (3) gives
\begin{eqnarray}
\partial_\mu B^\mu & = & \partial_\mu ( \bar \psi_R \gamma^\mu \psi_R +
\bar \psi_L \gamma^\mu \psi_L )=
\frac{1}{12\pi^2}
\partial^\mu \varepsilon_{\mu \nu \lambda \sigma} Sp ( A_R^\nu
A_R^\lambda A_R^\sigma ) = \nonumber \\
& = &  \frac{1}{24\pi^2}
\partial^\mu \varepsilon_{\mu \nu \lambda \sigma} Sp [ (U^+\partial^\nu U)
(U^+\partial^\lambda U)(U^+\partial^\sigma U) ] \; ,
\label{eq4} \end{eqnarray} \noindent
where we have introduced the meson field $U(x)=W^2(x)$ and the validity
of the last step follows after a little algebra.

Equation (4) suggests the following expression for the baryon number
current in terms of the meson field:
\begin{eqnarray}
B_\mu= \frac{1}{24\pi^2}
\varepsilon_{\mu \nu \lambda \sigma} Sp [ (U^+\partial^\nu U)
(U^+\partial^\lambda U)(U^+\partial^\sigma U) ] \; .
\label{eq5} \end{eqnarray}
Inserting here the Skyrme's hedgehog ansatz
\begin{eqnarray}
U(x)=\exp {\{i\frac{\vec{r}\cdot\vec{\tau}}{r}F(r)\}} \; ,
\label{eq6} \end{eqnarray} \noindent
with the boundary conditions
\begin{eqnarray}
F(0)=-n\pi \; , \; F(\infty)=0 \; ,
\label{eq7} \end{eqnarray} \noindent
we get after some computation (which is better to perform by some computer
program, for example be REDUCE \cite{17}):
\begin{eqnarray}
B&=&\frac{1}{24\pi^2}
\varepsilon_{i j k } \int Sp [ (U^+\partial_i U)
(U^+\partial_j U)(U^+\partial_k U) ]  d\vec{x} \nonumber \\
&=&\frac{1}{2\pi^2}\int\frac{\sin^2 F}{r^2}\frac{dF}{dr}d\vec{x}=
\frac{2}{\pi}\int_{F(0)}^{F(\infty)} \sin^2 F ~dF =n \; .
\nonumber \end{eqnarray}
So the hedgehog (6) with $n=1$ can be considered as a vortex with unit
baryon number, that is as a nucleon. But to make such an identification,
it should be shown that this hedgehog is a fermion, having a half-integer
spin.

Of course the half-integer spin is the most marvelous thing which can
be constructed from the spin zero pions. However this appears to be
a rather general phenomenon, not only a peculiarity of the Skyrmion.
In fact even a system from two spinless particles can have a half-integer
spin if one of them is such a queer object as a magnetic monopole.

The motion of a charge $e$ in the field of the magnetic monopole g is
described by the equation ($r^2=\vec{x}\cdot\vec{x}$)
\begin{eqnarray}
m\ddot{x_i}=\frac{eg}{4\pi}\frac{1}{r^3}\varepsilon_{i j k}x_j\dot{x_k}
\; . \nonumber \end{eqnarray}
From this we find easily
$$\frac{d}{dt}[\varepsilon_{i j k}x_j (m\dot{x_k})]=
-\frac{d}{dt}\left [\frac{eg}{4\pi}\frac{x_i}{r} \right ] \; . $$
Which indicates that for this system the angular momentum is \cite{18}
\begin{eqnarray}
J_i=\varepsilon_{i j k} x_j (m\dot{x_k})+\frac{eg}{4\pi}\frac{x_i}{r} \; .
\label{eq8} \end{eqnarray}
Therefore the system possesses a half-integer spin and is a fermion for
the lowest non-zero value $\frac{eg}{4\pi}=\frac{1}{2}$, allowed by the
Dirac's quantization condition \cite{19}.

This strange conclusion can be established on even more firm ground
by showing that (8) is the Neother current density corresponding
to an infinitesimal rotation.

As it is well known to find a singularity free Lagrangian for the
charge-monopole problem is not a trivial task. Nevertheless a very
elegant solution was found by Balachandran. By introducing
SU(2) matrix $s$, defined as $X=\sigma_i \hat{x}_i=s\sigma_3 s^+$,
instead of the angular variables $\hat{x}_i=x_i/r$, a non-singular
Lagrangian for the charge-monopole system can be written down in terms
of $s$ \cite{20,21}
\begin{eqnarray}
{\cal L}&=& \frac{1}{2}m \dot{x_i}\dot{x_i} +i\frac{eg}{4\pi}
Sp\{\sigma_3 s^+ \dot{s}\} = \nonumber \\
&=&\frac{1}{2}m\dot{r}^2+\frac{1}{4}mr^2 Sp\{\dot{X}^2\}+
i\frac{eg}{4\pi} Sp\{\sigma_3 s^+ \dot{s}\}  \; .
\label{eq9} \end{eqnarray}
Under a rotation we have $X^\prime=\exp{\{i\frac{\varepsilon_i}{2}\sigma_i\}}
X\exp{\{-i\frac{\varepsilon_i}{2}\sigma_i\}}$, that is $s^\prime =
\exp{\{i\frac
{\varepsilon_i}{2}\sigma_i\}}s$ and for an infinitesimal rotation
\begin{eqnarray}
\delta s=i\frac{\varepsilon_i}{2}\sigma_is \; , \;
\delta s^+=-i\frac{\varepsilon_i}{2}s^+\sigma_i \; , \;
\delta X=i\frac{\varepsilon_i}{2}[\sigma_i,X] \; .
\label{eq10} \end{eqnarray}
Using (10) we can find (note that $Sp \{ \dot{X}[\sigma_i,\dot{X}] \}=0$)
\begin{eqnarray}
\delta {\cal L}&=&\dot{\varepsilon}_i \left [ \frac{i}{4}mr^2Sp\{\dot{X}
[\sigma_i,X]\}-\frac{1}{2}\frac{eg}{4\pi}Sp\{X\sigma_i\}\right ] =
\nonumber \\
&=&-\dot{\varepsilon}_i\left [ \varepsilon_{i j k}x_j(m\dot{x}_k)+
\frac{eg}{4\pi}\hat{x}_i \right ] \; ,
\nonumber \end{eqnarray} \noindent
and so the corresponding Noether current $-\frac{\delta {\cal L}}{\delta
\dot{\varepsilon_i}}$ coincides indeed with (8).

But what has all this to do with Skyrmions? Among the high-derivative
terms, depicted by dots in (2), there is one which is analogous to
the charge-monopole interaction term. This so called Wess-Zumino term
has its root in the chiral anomaly \cite{22,23} and its contribution
to the action looks quite exotic
\begin{eqnarray}
&S_{WZ}=& \label{eq11} \\
&=\frac{-iN_c}{240\pi^2}\int d^5x ~\varepsilon_
{\alpha \beta \gamma \delta \sigma}
Sp\{(U^+\partial^\alpha U)(U^+\partial^\beta U)
(U^+\partial^\gamma U)(U^+\partial^\delta U)(U^+\partial^\sigma U) \}
\; . & \nonumber \end{eqnarray}
Like the charge-monopole system, it is not possible to write down
the corresponding piece of the (global) Lagrangian because the integral
in (11) is over a 5-dimensional disc whose boundary is 4-dimensional
Minkowskian space-time. But (11) simplifies enormously for the zero
modes of the Skyrmion which are a time dependent $SU(3)$ rotations $s(t)$
of the hedgehog $U(r)$:
$$ U(r,t)=s(t)U(r)s^+(t) \; . $$
The piece of (11) involving $s$ looks like \cite{23,24}
$$S_{WZ}=\frac{iN_cB}{2\sqrt{3}}\int dt ~Sp(\lambda_8 s^+ \dot{s}) $$
and exactly resembles the charge-monopole interaction term
$\sim Sp(\sigma_3 s^+ \dot{s})$. So the Wess-Zumino term produces
a "monopole in $SU(3)$ space" and can lead to a half-integer spin just
like how this happens in the charge-monopole system.

Note that the baryon number current (5) is just the Noether current
corresponding to the singlet vector symmetry $U \to \exp{(i\varepsilon
\frac{1}{N_c})}U\exp{(-i\varepsilon \frac{1}{N_c})}$ having the only
non-zero contribution from the Wess-Zumino term! This is not surprising
because both baryon number and the Wess-Zumino term have their origin
in the chiral anomaly, as was sketched above. More amusing is that
in the two flavor case the Wess-Zumino term vanishes while the baryon
number, like the cheshire cat's smile, still survives.

Thus the description of baryons as a mesonic vortexes is not so crazy
as looks at first sight. Moreover the idea proved his fruitfulness
in various applications \cite{25} and we expect that it will be useful
in the OZI rule violation studies also.

\section{Coherent state picture of the $N\bar N$ annihilation}
But first of all we need a description of the $N\bar N$ annihilation
at rest as a Skyrmion anti-Skyrmion annihilation. Numerical calculations
of this process have shown that just after the Skyrmion and anti-Skyrmion
touch classical pion wave emerges as a coherent burst and takes away
energy and baryon number as quickly as causality permits \cite{26}.
This observation led R.~Amado and collaborators to suggest the following
simplified version of $N\bar N$ annihilation at rest. After a very fast
annihilation a spherically symmetric "blob" of pionic matter of size
$\sim 1~fm$, baryon number zero and the total energy twice the nucleon
rest mass is formed. The further evolution of the system and the branching
rates of various channels are completely determined by the parameters
of this "blob" for which we can apply some simple phenomenological
parametrization, For example \cite{27}
\begin{eqnarray}
U(x)=\exp {\{i\frac{\vec{r}\cdot\vec{\tau}}{r}F(r)\}} \; , \;
F(r,t=0)=h\frac{r}{r^2+a^2}\exp {(-r/a)} \;
\label{eq12} \end{eqnarray}
where h is fixed by demanding that the total energy equals twice the
nucleon mass and $a$ is a range parameter, the only free parameter of
the model, assuming that the Skyrme model parameters are determined
from the static properties of the nucleon at their usual values.

Starting from the initial field configuration (12), we can use the
classical dynamical equations of motion to propagate the pion field,
and other fields coupled to it, far away from the annihilation region
there they no longer interact. These free classical radiation fields
should be appropriately quantized because we detect particles in the
final state and not the classical fields.

As was mentioned above numerical studies indicate that the resulting
lump of the pionic matter propagates after the annihilation as a coherent
burst of pion radiation. R.~Amado et al. suggested \cite{28} that this
peculiarity of the annihilation will be correctly represented if we
assume that the asymptotic quantum state in the radiation zone is in fact
a coherent state.

A coherent state $|\alpha>$ is defined as an eigenstate of the
annihilation operator \cite{29}
\begin{eqnarray}
a|\alpha>=\alpha |\alpha> \; .
\label{eq13} \end{eqnarray}
Using an usual commutation relation $[a,a^+]=1$ and representing
$|\alpha> = \sum_{n=0}^\infty c_n (a^+)^n|0>$, where $|0>$ is the vacuum
state, we find from (13) the recurrent relation $(n+1)c_{n+1}=
\alpha c_n$, from which it follows that up to normalization
$$ |\alpha>=\exp{(\alpha a^+)}|0> \;  . $$
It is not difficult to find out the normalization also. Let
$$<\alpha|\alpha> = <0|\exp{(\alpha^*a)}\exp{(\alpha a^+)}|0> =
f(\alpha^*,
\alpha).$$ It is clear that $f(0,\alpha)=f(\alpha^*,0)=1$. On the other
hand
\begin{eqnarray}
&\frac{\partial}{\partial \alpha}f(\alpha^*,\alpha)=& \nonumber \\
&=<0|\exp{(\alpha^*a)}a^+\exp{(\alpha a^+)}|0>=
<0|[\exp{(\alpha^*a)},a^+]\exp{(\alpha a^+)}|0>=& \nonumber \\
&=<0|(\frac{\partial}{\partial a}\exp{(\alpha^*a)})\exp{(\alpha a^+)}|0>
=\alpha^*f(\alpha^*,\alpha) \; , & \nonumber
\end{eqnarray}
so $f(\alpha^*,\alpha)=\exp{(\alpha^* \alpha)}$ and the normalized
coherent state looks like
\begin{eqnarray}
|\alpha>=\exp{(-\frac{\alpha^* \alpha}{2}) } \exp{(\alpha a^+)}|0> \;  .
\label{eq14} \end{eqnarray}

Let now $|f>$ be the coherent state corresponding to the positive energy
part of a quantum (scalar) field
$$\varphi(\vec{r},t)=\int \frac{d\vec{k}}{\sqrt{(2\pi)^32\omega_k}}
(a_ke^{i\vec{k}\cdot\vec{r}}e^{-i\omega_kt}+
a_k^+e^{-i\vec{k}\cdot\vec{r}}e^{i\omega_kt}) \; .$$
That is for all $\vec{k}$  we have $a_k|f>=f(\vec{k})|f>$. When
$$<f|\varphi(\vec{r},t)|f>=\int \frac{d\vec{k}}{\sqrt{(2\pi)^32\omega_k}}
(f(\vec{k})e^{i\vec{k}\cdot\vec{r}}e^{-i\omega_kt}+
f^*(\vec{k})f(\vec{k})e^{-i\vec{k}\cdot\vec{r}}e^{i\omega_kt}) $$
is the associated classical field. So the generalization of (14) will be
\begin{eqnarray}
|f>=\exp{\left(-\frac{1}{2}\int d\vec{k}f^*(\vec{k})f(\vec{k})+
\int d \vec{k} f(\vec{k})a^+_k\right)} |0> \; ,
\label{eq15} \end{eqnarray}
where $f(\vec{k})$ is the Fourier transform of the classical radiation
field.

The coherent state (15) has no definite 4-momentum. For pions it is also
necessary to have a coherent state with a definite isospin. How to handle
these subtleties the reader can find in the original literature
\cite{28,30,31}.

Thus the logical scheme of the R.~Amado et al.'s approach to the $N\bar N$
annihilation at rest looks like this \cite{27}:

Given the initial (pion) field configuration and using the classical
dynamical equations the asymptotic $\pi, \rho, \omega \dots$ fields
should be generated. From those fields we construct one corresponding
coherent state with definite four-momentum and isospin. To find
the probability of some final state we have to calculate the overlap of
this state to the coherent state, which gives us the respective transition
amplitude. The probability when is calculated as usual by integrating
the absolute square of this transition amplitude over all final particle
momenta and summing over all possible values of intermediate $\pi$ and
$\rho$ isospin.

The physical reason why the above given picture of the $N \bar N$
annihilation is reasonable lies in the large number of produced particles.
For the large final pion number its field can be approximated as classical.
On the other hand if the number of quanta is very large an elimination
of one of them doesn't make a big difference. So the corresponding quantum
state is in a good approximation an eigenstate of the annihilation operator.

Besides coherent states provide the closest quantum analog to the classical
dynamics in the following sense. If we consider the standard position and
momentum operators
$$q=\sqrt{\frac{\hbar}{2\omega}}(a^++a) \; \; , \; \;
p=i\sqrt{\frac{\hbar\omega}{2}}(a^+-a)  \; , $$
when in a coherent state $|\alpha>$ we have the corresponding uncertainties
$$(\Delta q)^2=<\alpha|q^2|\alpha>-<\alpha|q|\alpha>^2=
\frac{\hbar}{2\omega} \{ [1+(\alpha+\alpha^*)^2]-(\alpha+\alpha^*)^2 \}=
\frac{\hbar}{2\omega} \; , $$
$$(\Delta p)^2=<\alpha|p^2|\alpha>-<\alpha|p|\alpha>^2=
-\frac{\hbar\omega}{2} \{ [(\alpha^*-\alpha)^2-1]-(\alpha^*-\alpha)^2 \}=
\frac{\hbar\omega}{2} \; . $$
So the celebrated uncertainty relation is saturated for the coherent states
\cite{29}
$$(\Delta q)^2(\Delta p)^2=\frac{\hbar^2}{4} \; .$$

In reality, however, the average number of pions produced in the $N \bar N$
annihilation at rest is about five with a variance of one, and so it is not
a very large. Nevertheless a remarkably good agreement between calculated
and measured characteristics of the annihilation was found by R.~Amado et al.
\cite{27,28,30,31}. Note that the omega \cite{30} and rho \cite{31}
mesons were also successfully included into the model despite the fact that
their average numbers $\sim 1$.

\section{The OZI rule violation}
It is easy to see that the above described R.~Amado et al.'s model of
$N \bar N$ annihilation naturally incorporates the OZI rule violation for
the $\phi$-meson production.

In the case of the ideal $\omega - \phi$ mixing, $\phi$-meson field
decouples from the Skyrmion and is not directly excited by rotating hedgehog
\cite{32}, which is believed to represent a nucleon in the collective
coordinate approach \cite{23,25}. On the contrary $K$-meson field appears
in the action with a linear coupling to the time derivative of the rigidly
rotating hedgehog, because of the peculiar properties of the Wess-Zumino
term and becomes excited \cite{33}.

So a nucleon in the Skyrme model carries some amount of the $K - \bar K$
cloud around it -- the analog of the supposed strange component of the
nucleon wave function, mentioned in the introduction. This means that after
the annihilation initial field configuration contains not only pion field
but also kaon (and antikaon) field(s). According to the spirit of the
R.~Amado et al.'s approach we have to evolve this kaon field using the
classical equations of motion in the radiation zone and include in the
coherent state. When from this coherent state we can extract the
annihilation branching ratios for the kaon containing channels. But more
important for us now is the excitation of the $\phi$-meson field due to
its dynamical coupling to the $K$-meson field. While kaon field develops
from the annihilation point to the radiation zone, $\phi$-meson field
inevitably emerges because of obvious $K \bar K - \phi$ coupling and the
coherent state describing annihilation should contain $\phi$-meson also,
despite the fact that it is absent in the initial lump of the Skyrmion
matter!

This transformation of the initial strangeness into the final $\phi$-meson
resembles the rescattering model discussed in the introduction. Thus the
Skyrme model picture of the $N \bar N$ annihilation at rest not only gives
a natural framework for the OZI rule violation in the $\phi$-meson
production, but also incorporates the characteristic features of two, at
first glance very different, main models suggested to explain this
phenomenon!

It is amusing that the  OZI rule violation in the $\phi$-meson
production appears to be one more manifestation of the Wess-Zumino term,
because this very term is responsible for the initial kaon excitation
in the nucleon which afterwards transforms into the $\phi$-meson.

\end{document}